\documentclass[a4paper,12pt,twoside]{article}

\usepackage[T1]{fontenc}
\usepackage[utf8]{inputenc}

\usepackage{graphicx}
\usepackage{grffile}  
\usepackage{array}
\usepackage{multirow}
\usepackage{float}

\usepackage[left=2cm,right=2cm,top=2cm,bottom=4cm]{geometry}

\usepackage[
linktocpage=true,
  colorlinks   = true,    
  urlcolor     = blue,    
  linkcolor    = blue,    
  citecolor    = red      
]{hyperref}
\usepackage{orcidlink} 

\usepackage{amsmath}

\usepackage{xspace}

\usepackage{color}

\usepackage[
backend=biber,
bibencoding=utf8,
style=numeric-comp,
sorting=none
]{biblatex}
\renewbibmacro{in:}{}

\usepackage[affil-it]{authblk}

\author[1]{Jan Kalinowski\orcidlink{0000-0001-5618-0141}}
\author[2]{Wojciech Kotlarski\orcidlink{0000-0002-1191-6343}}
\author[1]{Krzysztof Mękała\orcidlink{0000-0003-4268-508X}}
\author[1]{Kamil Zembaczynski}
\author[1*]{Aleksander Filip Żarnecki\orcidlink{0000-0001-8975-9483}}

\title{New approach to DM searches with mono-photon signature}

\affil[1]{Faculty of Physics, University of Warsaw, Poland}
\affil[2]{National Centre for Nuclear Research, Warsaw, Poland}
\affil[*]{filip.zarnecki@fuw.edu.pl}

\newcommand{\whizard}{\textsc{Whizard}\xspace}

\newcommand{\delphes}{\textsc{Delphes}\xspace}
\newcommand{\epem}{\ensuremath{\textrm{e}^+\textrm{e}^-}\xspace}

\DeclareMathAlphabet{\mathcal}{OMS}{cmsy}{m}{n}
\newcommand{\kkmc}{${\cal KK}$\,MC\xspace}

\graphicspath{ {./plots/} }

\batchmode

\begin{document}

\maketitle

\vspace{2cm}

\begin{abstract}

High energy \epem colliders offer unique possibility for the most
general dark matter search based on the mono-photon signature.
Analysis of the energy spectrum and angular distributions of 
photons from the initial state radiation can be used 
to search for hard processes with invisible final state
production.

Most studies in the past focused on scenarios assuming heavy mediator
exchange.
We notice however, that scenarios with light mediator exchange are
still not excluded by existing experimental data, if the mediator
coupling to Standard Model particles is very small.
We  proposed a novel approach, where the experimental sensitivity to
light mediator production is defined in terms of both the mediator
mass and mediator width.
This approach is more model independent than the approach assuming
given mediator coupling values to SM and DM particles.  

Summarised in this contribution are published results of our studies
concerning simulation of mono-photon events with \whizard and the 
expected sensitivity of the International Linear Collider (ILC) and
Compact Linear Collider (CLIC) experiments to dark matter production.

\end{abstract}

\newpage

\section{Introduction}

\begin{figure}[htb]
  \centerline{\includegraphics[width=0.3\textwidth]{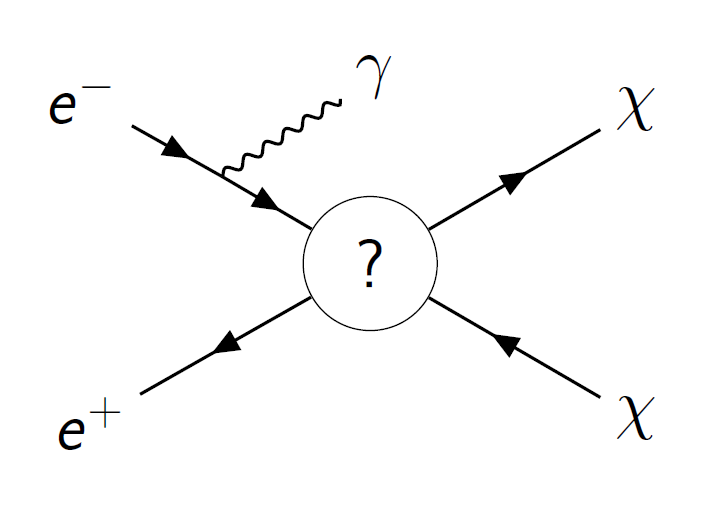}}
  \caption{Diagram describing DM particle pair production process
    with additional ISR photon radiation.}
  \label{fig:sigdiag}
\end{figure}

Direct pair-production of DM particles can be searched for  at high
energy \epem colliders.
This process can be detected, if additional hard photon radiated from
the initial state, see Fig.~\ref{fig:sigdiag}, is observed in the detector.
This so called mono-photon signature is considered as the most general
approach to search for DM particle production.

Presented in this contribution are results concerning the DM pair
production with mono-photon signature at future linear \epem
colliders, ILC \cite{Adolphsen:2013kya} and CLIC \cite{Robson:2018enq}.
Baseline ILC design assumes initial stage at 250\,GeV, followed by
500\,GeV and 1\,TeV as the possible upgrade \cite{Bambade:2019fyw}.
Polarisation is assumed for both e$^-$ and e$^+$ beams, of 80\% and
30\%, respectively. 
Total of 4000\,fb$^{-1}$ of data is expected to be collected at
500\,GeV stage, with 80\% of the integrated luminosity taken with LR
and RL beam polarisation combinations (2$\times$1600\,fb$^{-1}$), and
only 20\% with RR and LL beam polarisation combinations
(2$\times$400\,fb$^{-1}$).
Novel two-beam acceleration scheme proposed for CLIC opens the
possibility of reaching the collision energy of up to 3\,TeV. 
Total integrated luminosity of 5000\,fb$^{-1}$ is
expected at 3\,TeV stage, with 80\% (4000\,fb$^{-1}$) collected
with left-handed electron beam polarisation and 20\% (4000\,fb$^{-1}$)
with right-handed electron beam \cite{CLICdp:2018cto}.
Positron beam polarisation is not included in the CLIC baseline design.

\section{Simulating mono-photon events}

Precise and consistent simulation of BSM processes and of the SM
backgrounds is crucial for proper estimate of the experimental
sensitivity to processes with mono-photon signature.
Procedure developed for simulating these processes with \whizard
\cite{Moretti:2001zz,Kilian:2007gr} is described in a dedicated paper
\cite{Kalinowski:2020lhp}. 
We summarise our main results below.

\whizard program, which is widely
used for \epem collider studies, provides the ISR structure function
option that includes all orders of soft and soft-collinear photons as
well as up to the third order in high-energy collinear photons. 
However, photons generated by \whizard in this approximation can not
be considered as ordinary final state particles, as they represent all
photons radiated in the event from a given lepton line.
Nor the ISR structure function can properly account for hard non-collinear
photon radiation.
The proper solution is to generate all ``detectable'' photons on the 
Matrix Element (ME) level.
This however requires a proper procedure for matching the soft ISR
radiation with the hard ME simulation, to avoid double-counting. 

\begin{figure}[tb]
  \centerline{
  \includegraphics[width=0.4\textwidth]{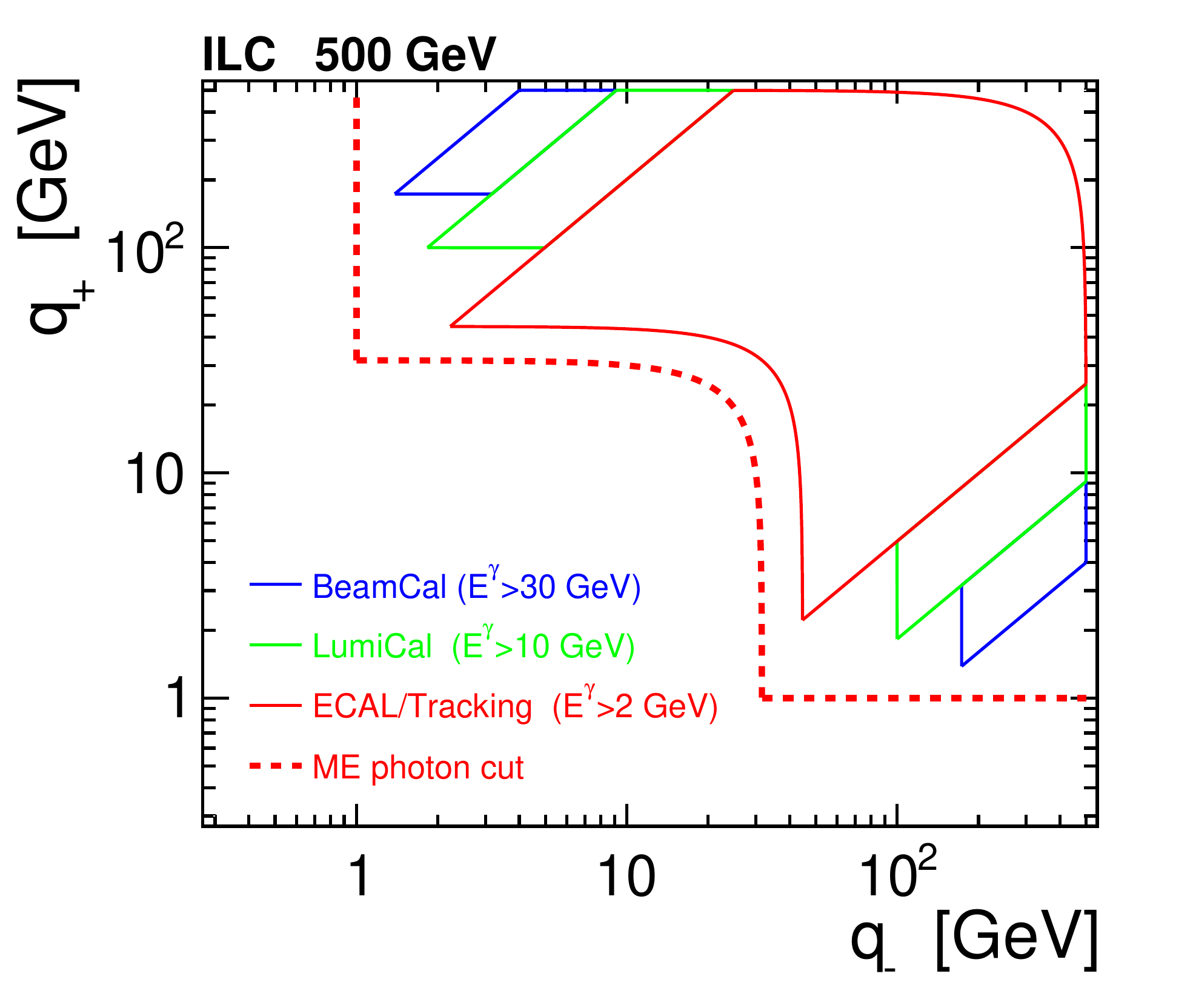}
  \includegraphics[width=0.4\textwidth]{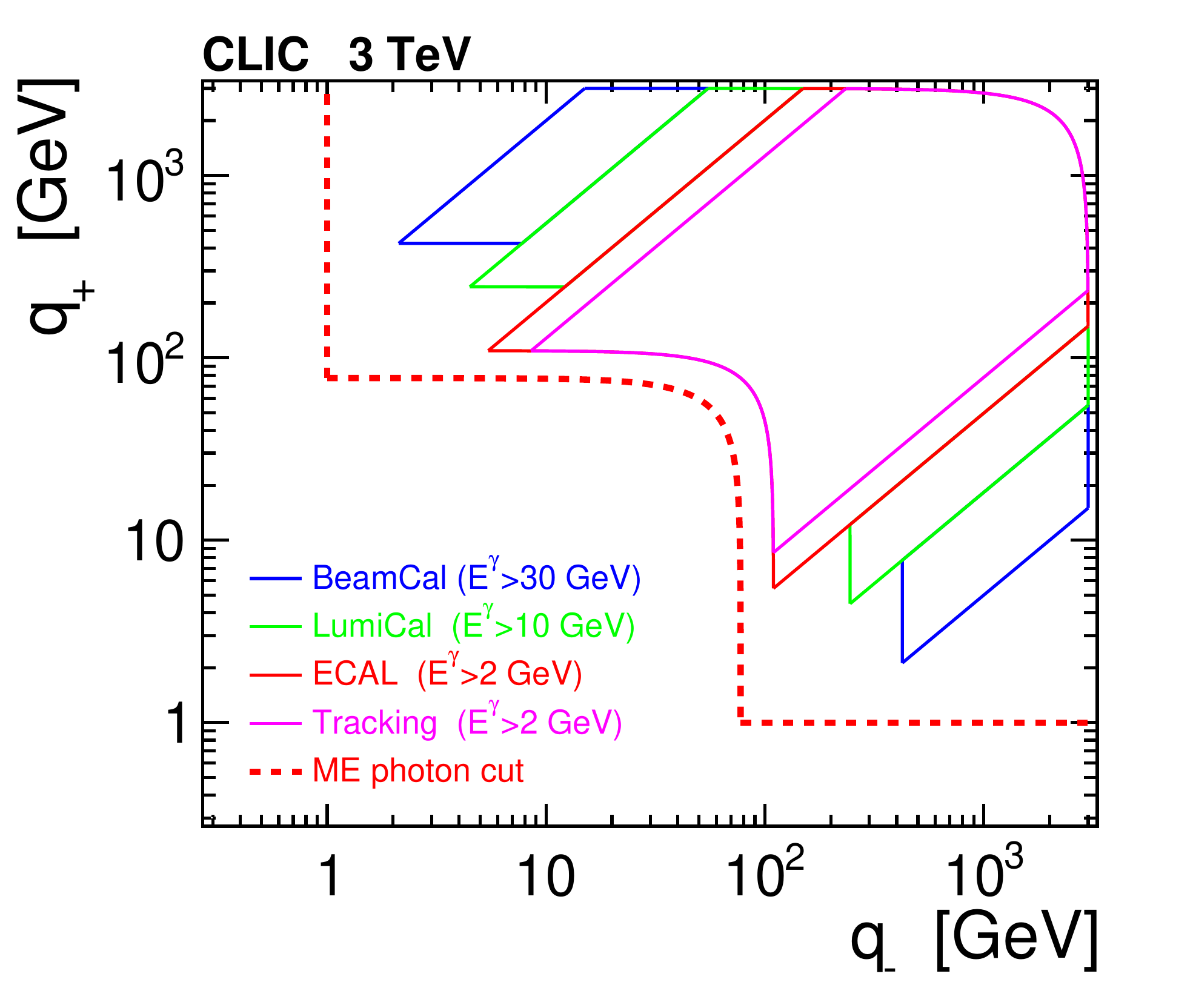}
  }
  \caption{Detector acceptance in the $(q_+,q_-)$ plane expected for
    the future experiments at 500\,GeV ILC (left) and 3\,TeV CLIC
    (right). Red dashed lines indicate the cut
    used to restrict the phase space for ME photon generation
    \cite{Kalinowski:2021tyr}. 
   }
  \label{fig:q_plot}
\end{figure}

The procedure for matching ISR and ME regimes proposed
in \cite{Kalinowski:2020lhp}
is based on two variables, calculated separately for each emitted
photon, used to describe kinematics of the photon emission:
\begin{eqnarray*}
  q_{-} & = & \sqrt{4 E_{_0} E_\gamma} \cdot
  \sin{\frac{\theta_\gamma}{2}} \; , \\
  q_{+} & = & \sqrt{4 E_{_0} E_\gamma} \cdot
  \cos{\frac{\theta_\gamma}{2}} \; ,
\end{eqnarray*}
where $E_{_0}$ is the nominal electron or positron beam energy, while $E_\gamma$
and $\theta_\gamma$ are the energy and scattering angle of the emitted
photon in question.
The detector acceptance in the $(q_+,q_-)$ plane expected for the future
ILC and CLIC experiments is presented in Fig.~\ref{fig:q_plot}.
Red dashed lines indicating the cut used to separate ``soft ISR''
emission region (to the left and below the dashed line) from the
region described by ME calculations (to the right and above the dashed
line) shows that with this procedure only the photons generated on the ME
level can enter the detector acceptance region.
Validity of the proposed matching procedure was verified by comparing
results of the \whizard simulation with those from the semi-analytical
\kkmc code \cite{Jadach:1999vf,Jadach:2013aha}, for the radiative neutrino pair-production events.
Details can be found in \cite{Kalinowski:2020lhp}.

Results concerning sensitivity of future linear \epem colliders to
processes of dark matter production with light mediator exchange were
presented in \cite{Kalinowski:2021tyr}.
Dedicated model \cite{SimpDMwiki} was encoded into \textsc{FeynRules}
\cite{Christensen:2008py,Alloul:2013bka} 
for calculating the DM pair-production cross section and generating
signal event samples with \whizard.  
We consider mediator mass, width and coupling to electrons as the
independent model parameters, with the total mediator width assumed to
be dominated by its decay to the DM particles. 
In this approximation, cross section dependence on the DM particle
couplings is absorbed in the total mediator width and the results
hardly depend on the DM particle type or coupling structure.

\begin{figure}[tb]
  \centerline{
 \includegraphics[width=0.4\textwidth]{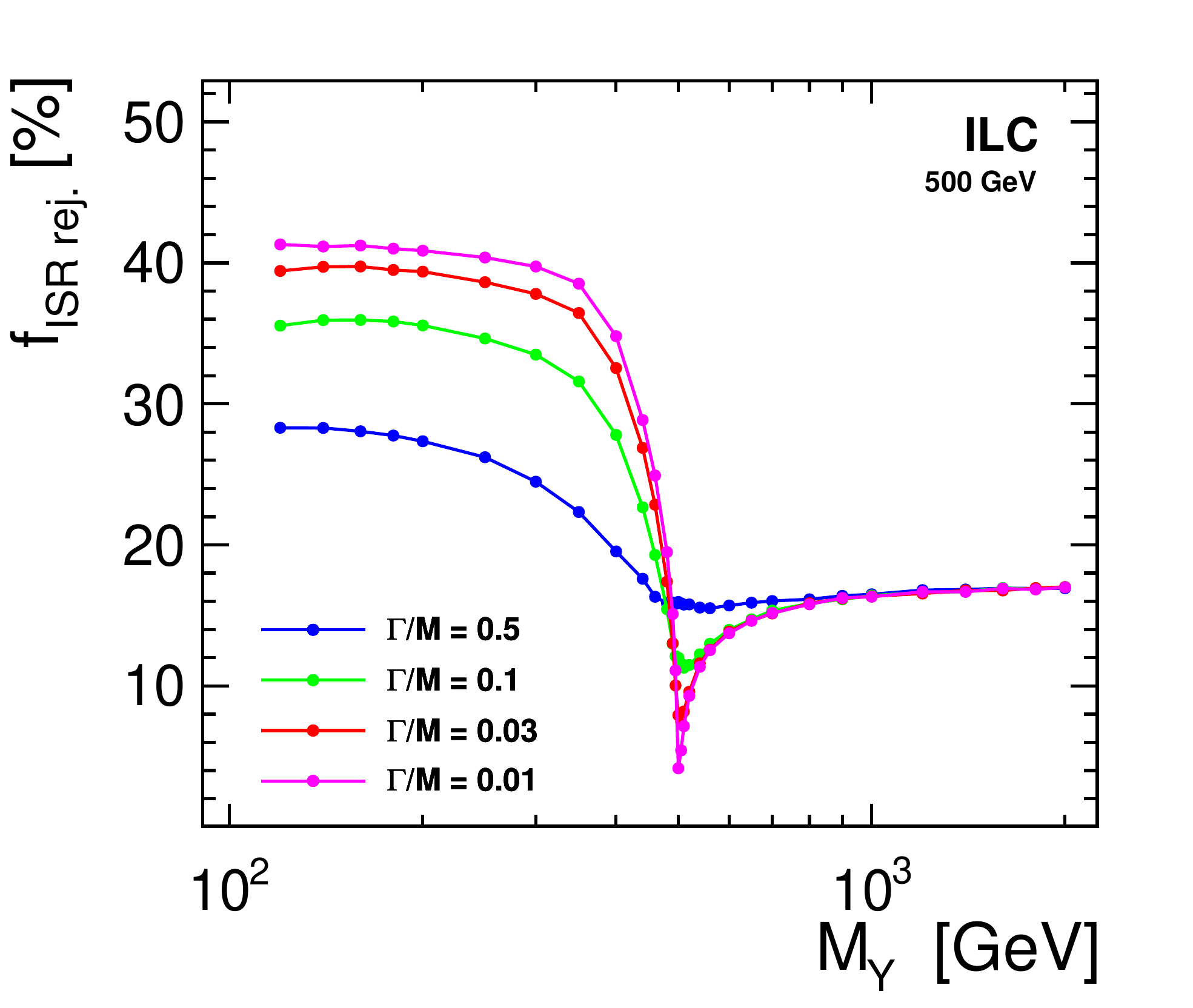}
 \includegraphics[width=0.4\textwidth]{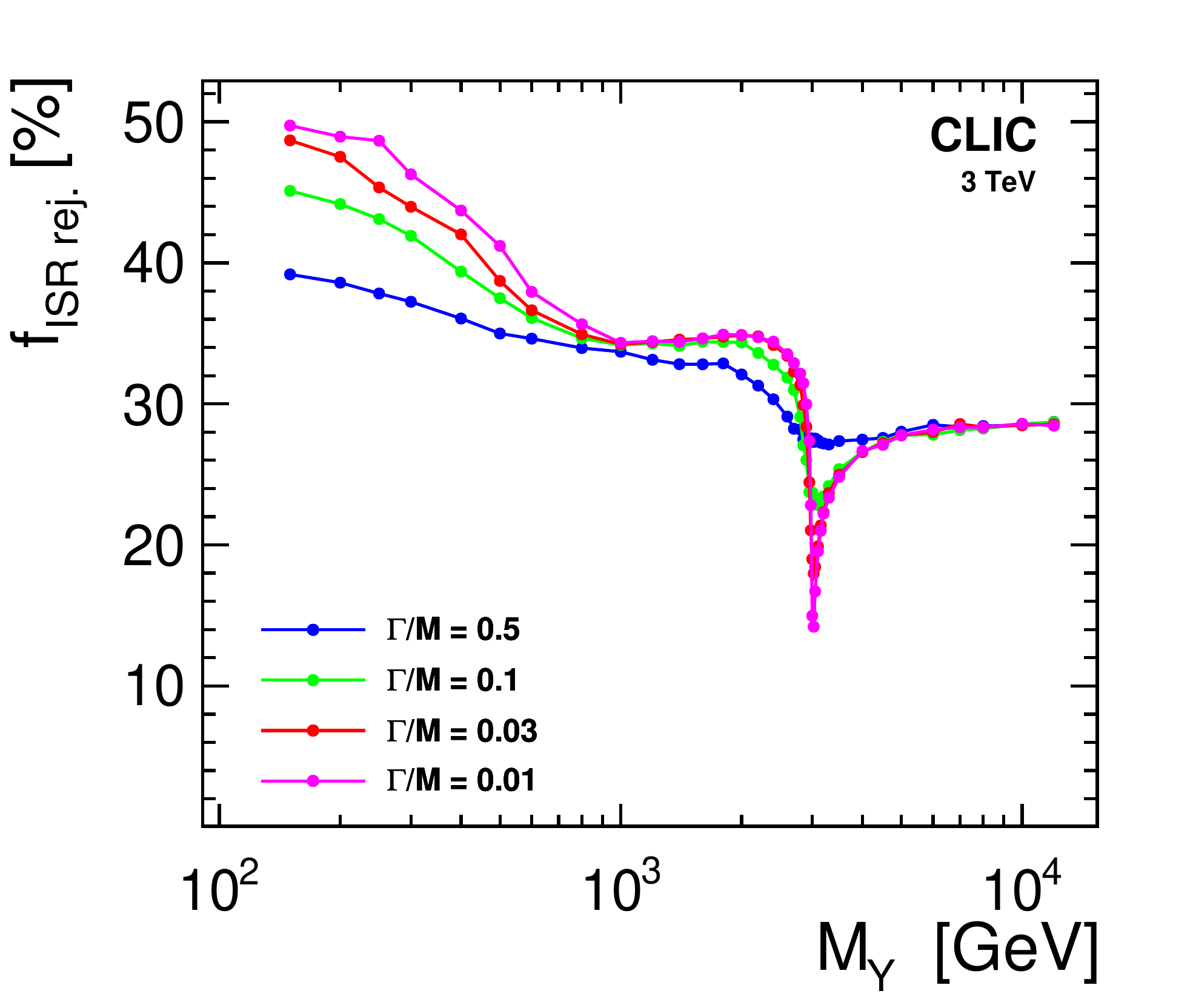}
 }
  \caption{Fraction of \whizard events, which are removed by the ISR
    rejection procedure, as described in
    \cite{Kalinowski:2020lhp}. Figure taken from
    \cite{Kalinowski:2021tyr}.
  }
  \label{fig:isr_rej}
\end{figure}

\begin{figure}[tb]
  \centerline{
 \includegraphics[width=0.4\textwidth]{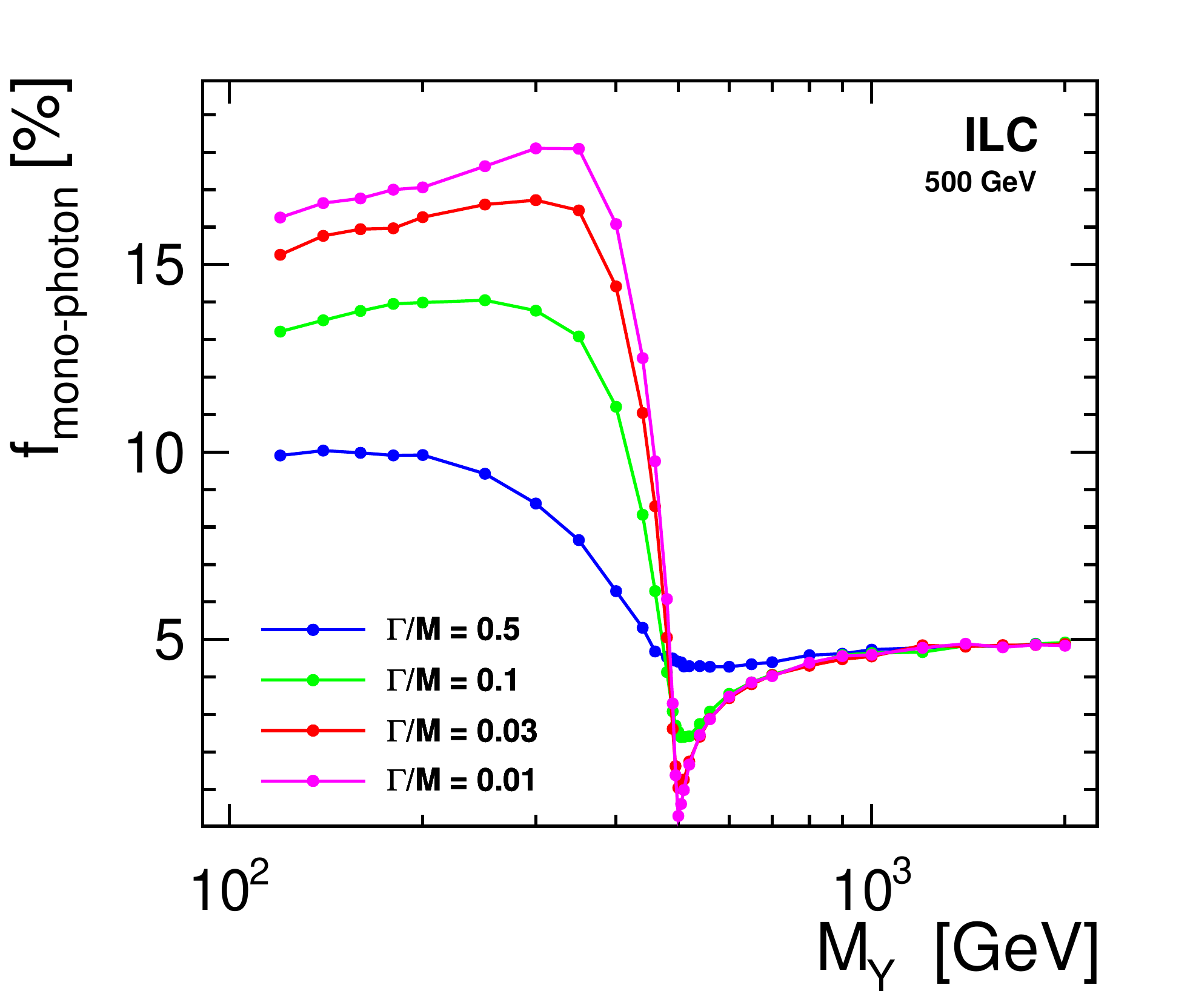}
 \includegraphics[width=0.4\textwidth]{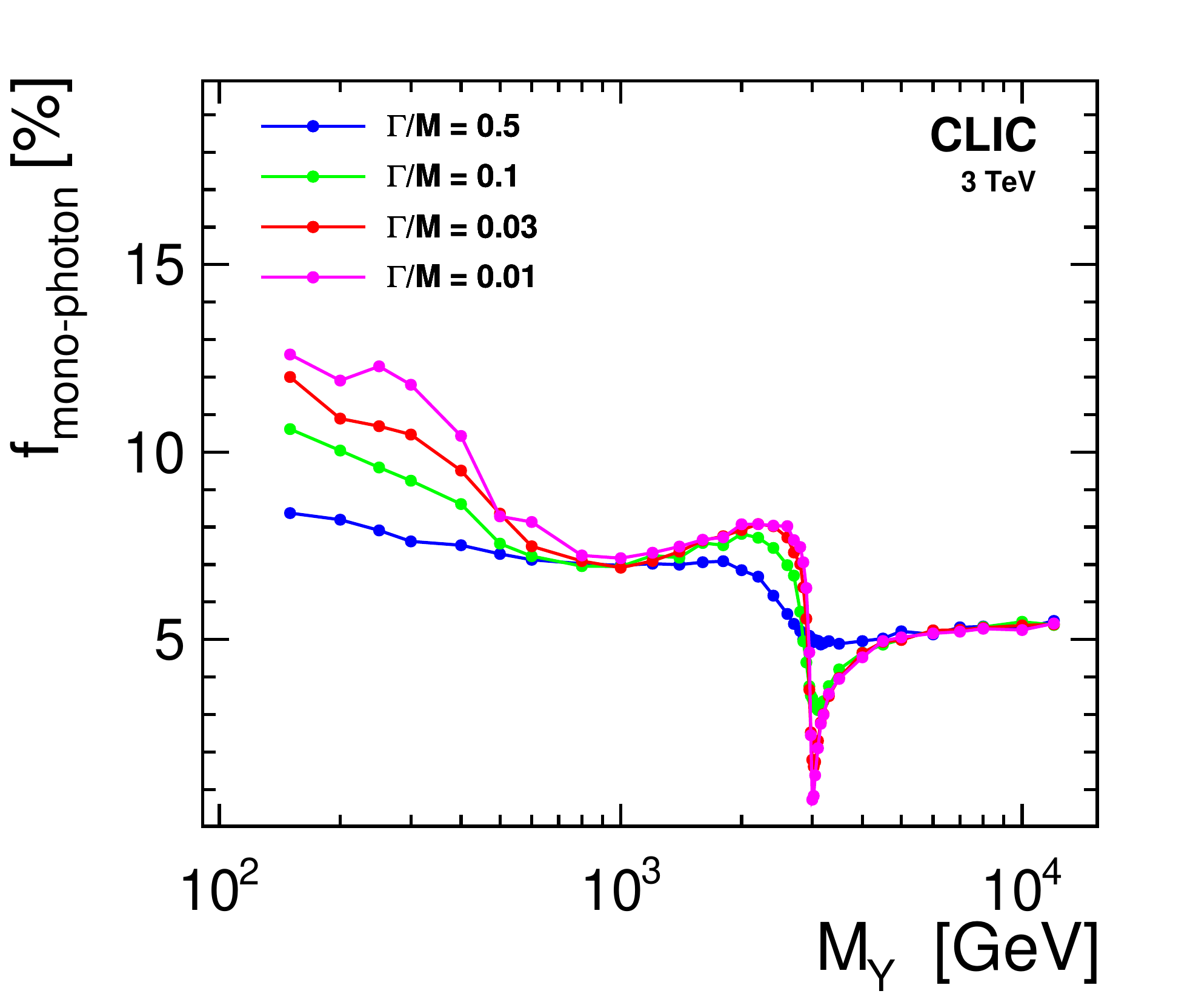}
 }
  \caption{Fraction of dark matter pair-production events, which are
    reconstructed as mono-photon events in the detector, as a function
    of the assumed mediator mass, for the ILC running at 500\,GeV
    (left) and CLIC running at 3\,TeV (right) and different fractional
    mediator widths, as indicated in the plot.
  }
  \label{fig:det_effi}
\end{figure}

The matching procedure described in \cite{Kalinowski:2020lhp}, removing 
events with ISR photons emitted in the ME phase space region
(so called ``ISR rejection'')
can result in up to 50\% correction to the DM production cross
section, as shown in Fig.~\ref{fig:isr_rej}.
Most of the DM pair-production events will remain ``invisible'' in the
detector.
While radiation of one or more photons (on the ME level) is expected in up
to 50\% of these events, most of these photons go along the beam line
and only a small fraction is reconstructed as mono-photon events in
the detector. 
The fraction of ``tagged'' events also depends significantly on the
mediator mass and width, as shown in Fig.~\ref{fig:det_effi}.
Presented results are based on the fast detector simulation framework
\delphes \cite{delph} in which the two detector models were implemented,
including detailed description of the calorimeter systems in the very
forward region.

\section{Sensitivity to dark matter production}

\begin{figure}[tb]
  \centerline{%
\includegraphics[width=0.4\textwidth]{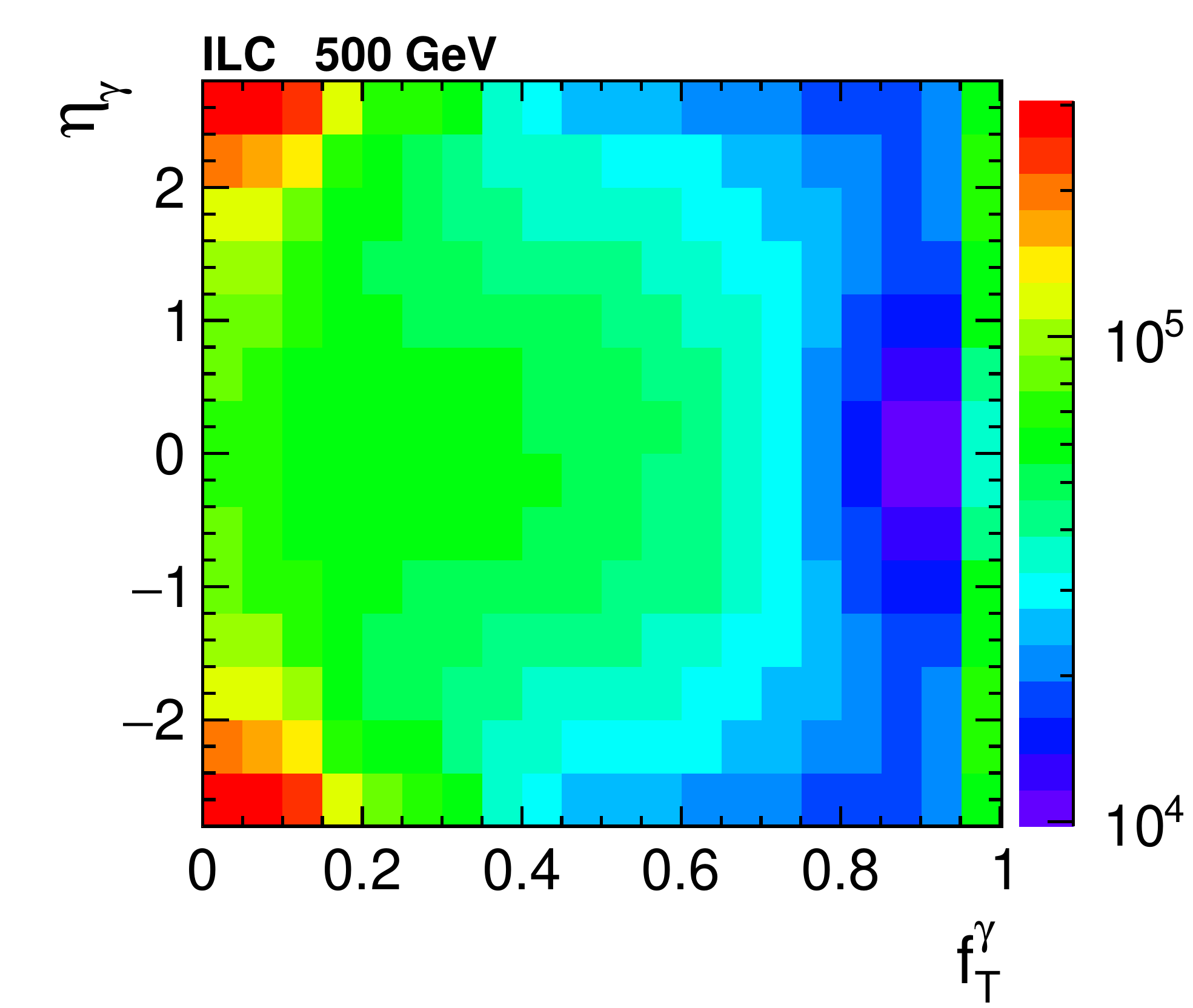}%
\includegraphics[width=0.4\textwidth]{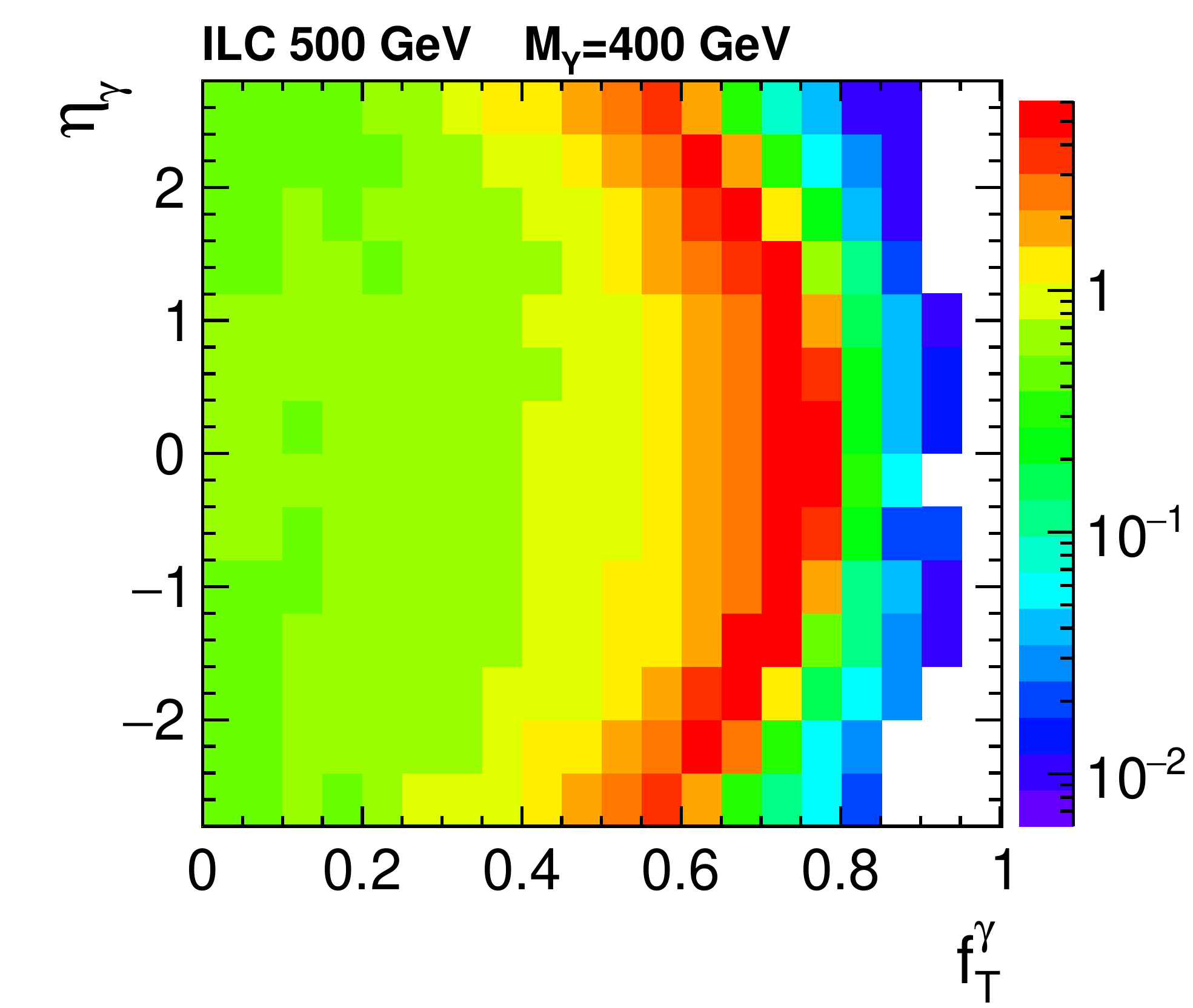}%
}
\caption{Pseudorapidity vs transverse momentum fraction for mono-photon
events at 500\,GeV ILC running with –80\%/+30\% electron/positron beam
polarisation and integrated luminosity of 1.6\,ab$^{-1}$.
Left: for sum of considered SM backgrounds. Right: for
pair-production of Dirac fermion DM particles with m$_\chi$ = 50\,GeV
and vector mediator mass of M$_Y$ = 400\,GeV, assuming total
production cross section of 1\,fb \cite{Kalinowski:2021tyr}.
  \label{fig:2d}
}
\end{figure}

Summarised below are results of \cite{Kalinowski:2021tyr}
addressing pair-production of DM particles at the ILC and
CLIC for scenarios with both light and heavy mediators.
Scenarios with light mediator exchange are still not
excluded by the existing experimental data.
Limits on the mediator coupling to electrons which were set at LEP and
by the LHC experiments, are of the order of 0.01 or above. 
The study focused on scenarios with very small mediator couplings to
SM, when the total mediator width is dominated by invisible decays,
$\Gamma_\mathrm{SM} \ll \Gamma_\mathrm{DM} \approx \Gamma_\mathrm{tot}$.
``Experimental-like'' approach is adopted, focused on setting the DM
pair-production cross section limits as a function of the mediator mass
and width, assuming DM particles are light (the mass of fermionic DM is
fixed to m$_\chi$ = 50\,GeV for all results presented in the following).
Limits on the production cross section are extracted from the
two-dimensional distributions of the reconstructed mono-photon
events in pseudorapidity and transverse momentum fraction.
Distributions expected at 500\,GeV ILC, for SM
backgrounds and example DM production scenario, are compared in
Fig.~\ref{fig:2d}. 
Transverse momentum fraction, f$^\gamma_{\mathrm{T}}$, is a logarithm of
the transverse momentum scaled to span the range between the minimum
and maximum photon transverse momentum allowed for given rapidity.

\begin{figure}[tbp]
\centerline{%
 \includegraphics[width=0.4\textwidth]{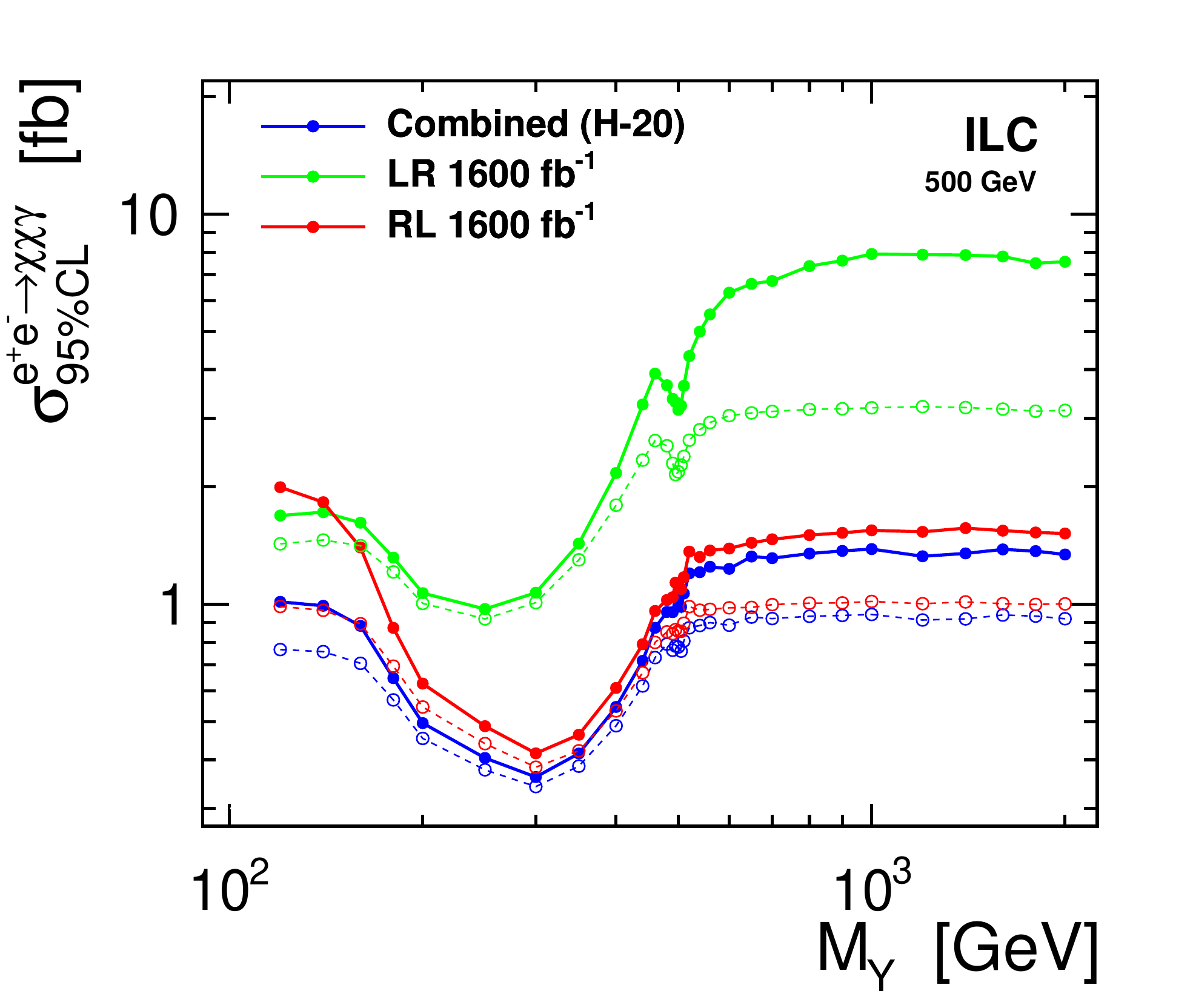}%
 \includegraphics[width=0.4\textwidth]{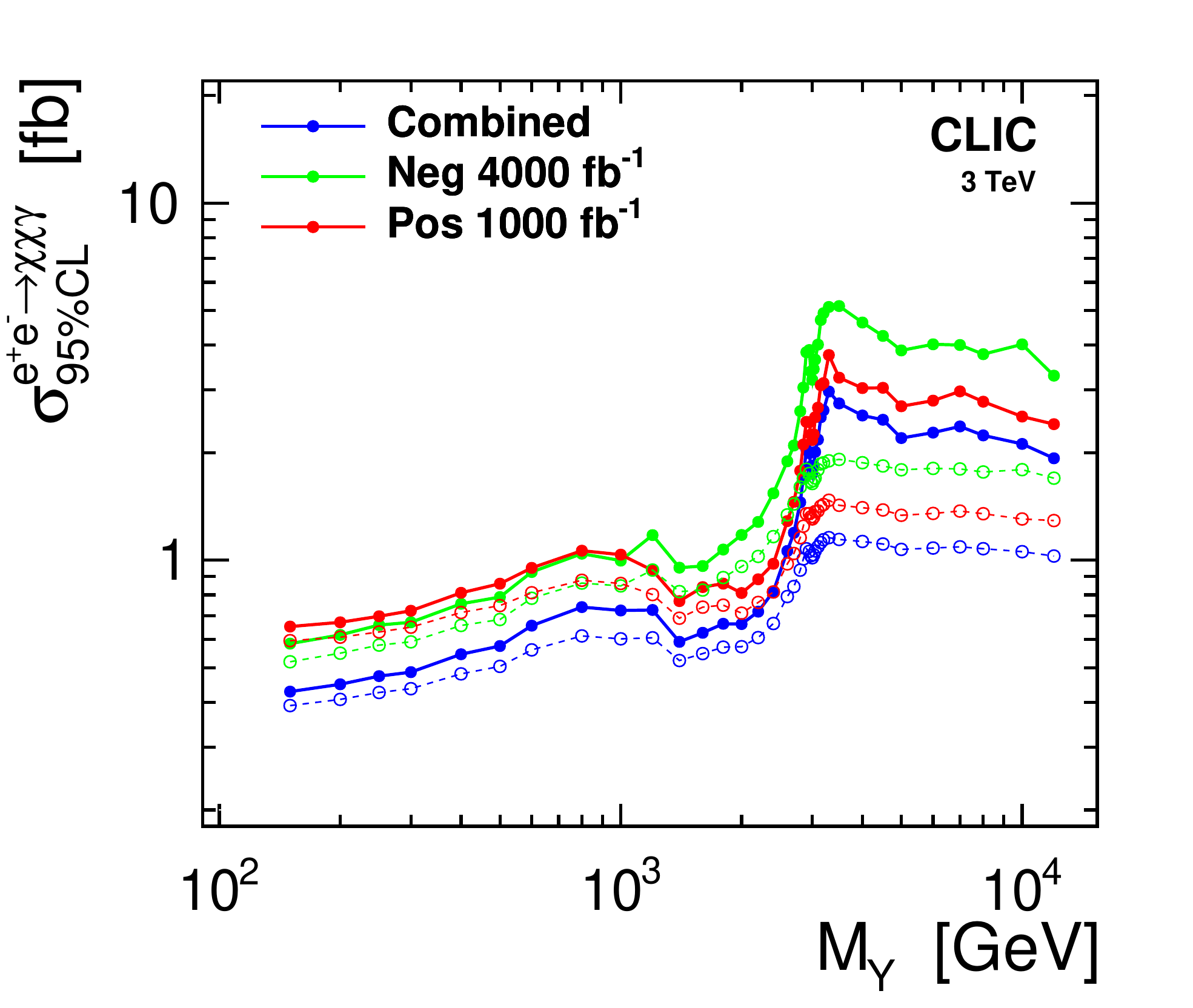}%
 }
  \caption{Limits on the cross section for the radiative light DM
    pair-production processes with vector mediator exchange at
    500\,GeV ILC (left) and 3\,TeV CLIC (right), for mediator 
    width $\Gamma/M = 0.03$, with (solid line) and without (dashed
    line) taking into account systematic uncertainties
    \cite{Kalinowski:2021tyr}. 
  } 
  \label{fig:res_sys}
\end{figure}

Cross section limits for radiative DM production (for events with
the tagged photon) at 500\,GeV ILC and 3\,TeV CLIC, for vector
mediator exchange scenario, are compared in Fig.~\ref{fig:res_sys}.
Combined analysis of data taken with different beam polarisation
combinations results in strongest limits, also reducing the impact of
systematic uncertainties.
Systematic effects are also suppressed when searching for on-shell 
production of narrow mediator, i.e. for M$_\mathrm{Y} < \sqrt{s}$
(assuming $\Gamma/\mathrm{M} \ll 1$).

\begin{figure}[tbp]
\centerline{%
  \includegraphics[width=0.4\textwidth]{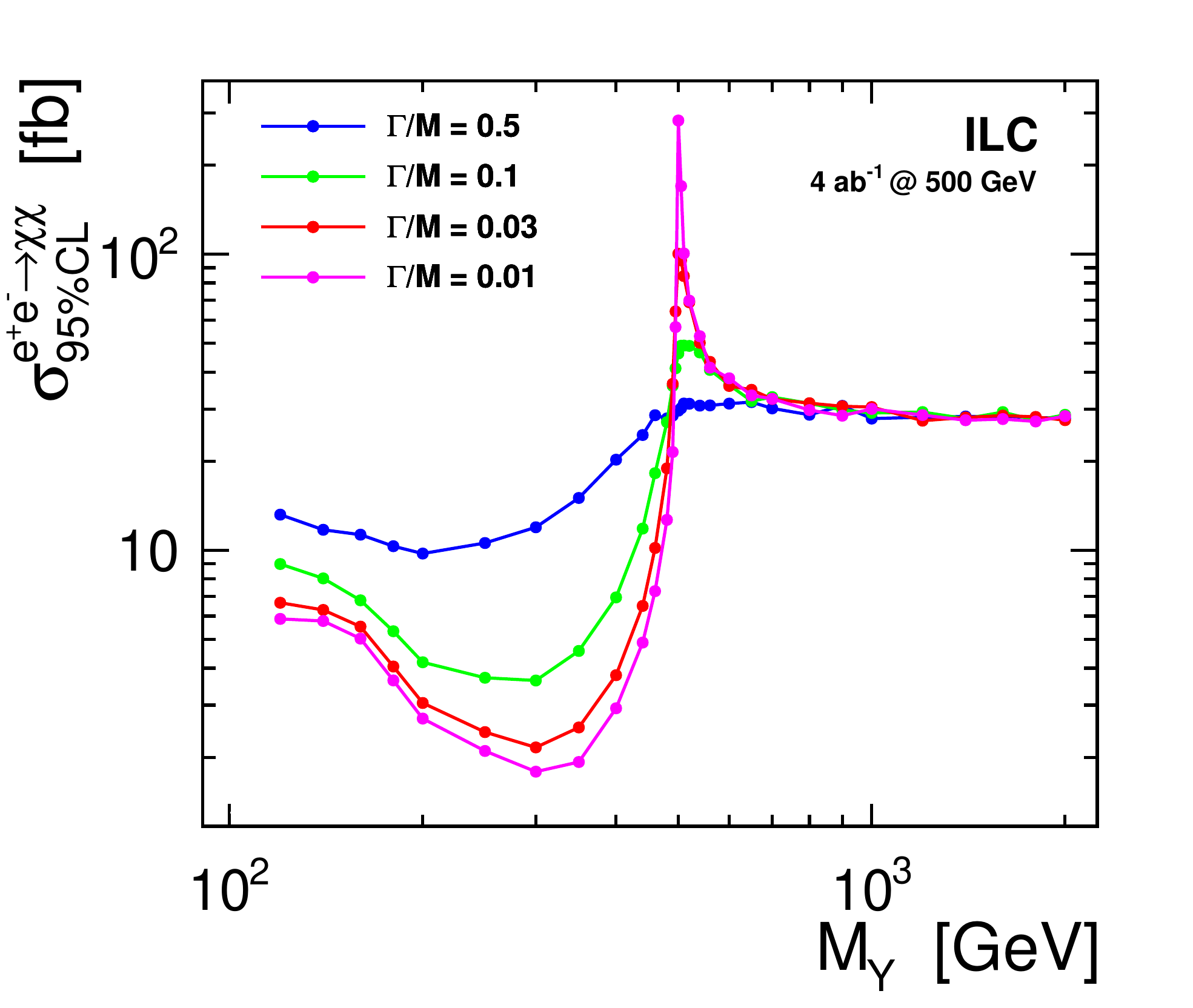}%
  \includegraphics[width=0.4\textwidth]{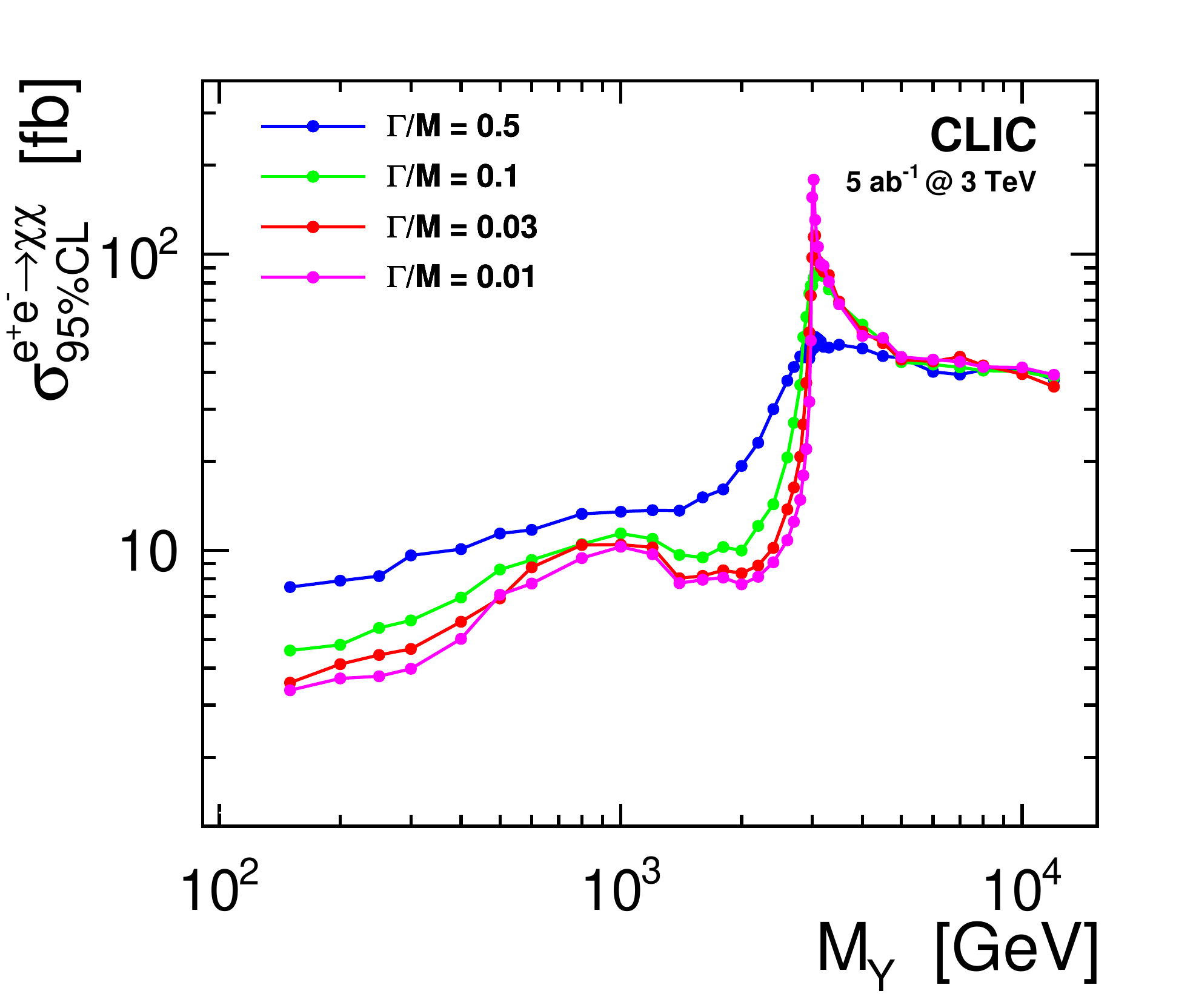}%
}
\caption{Limits on the cross section for light fermionic DM
pair-production processes with $s$-channel mediator exchange
for the ILC running at 500\,GeV (left) and CLIC running at 3\,TeV
(right), for the vector mediator exchange and different fractional
mediator widths.
Combined limits corresponding to the assumed running scenarios
are presented with systematic uncertainties taken into account
\cite{Kalinowski:2021tyr}. 
} 
  \label{fig:res_wid}
\end{figure}

After correcting for the hard photon tagging probability (refer
Fig.~\ref{fig:det_effi}), limits for the total DM pair-production
cross section can be extracted.
Presented in Fig.~\ref{fig:res_wid} are limits expected from the
combined analysis of data taken with different beam polarisations,
for different fractional mediator widths assuming vector mediator
exchange.
Strongest limits are obtained for processes with light mediator
exchange and for narrow mediator widths, whereas for heavy mediator
exchange (M$_\mathrm{Y} \gg \sqrt{s}$) cross section limits no longer
depend on the mediator width.
Limits are significantly weaker for narrow mediator with M$_\mathrm{Y}
\approx \sqrt{s}$, when photon radiation is significantly suppressed.

\begin{figure}[tb]
\centerline{%
 \includegraphics[width=0.4\textwidth]{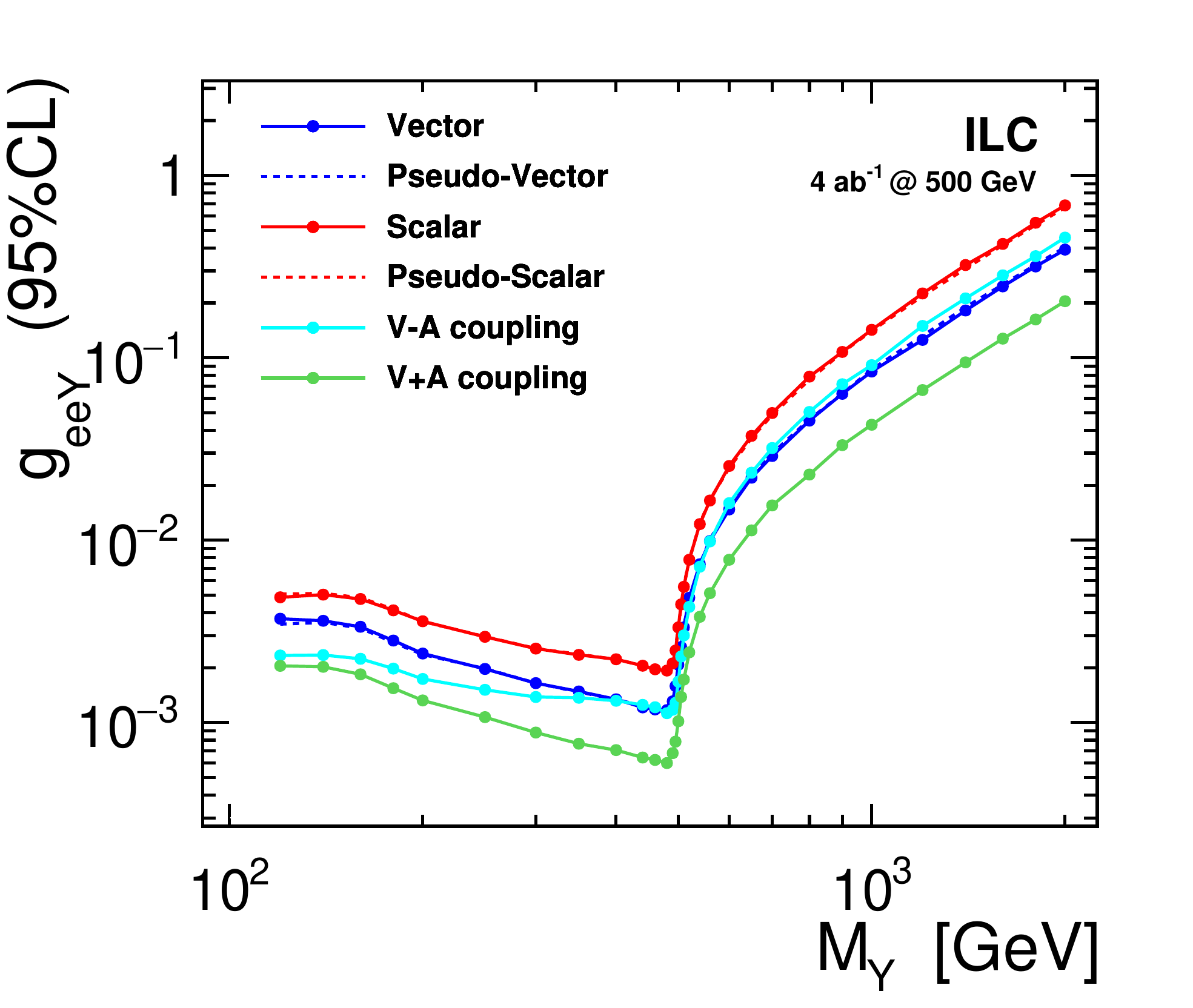}%
 \includegraphics[width=0.4\textwidth]{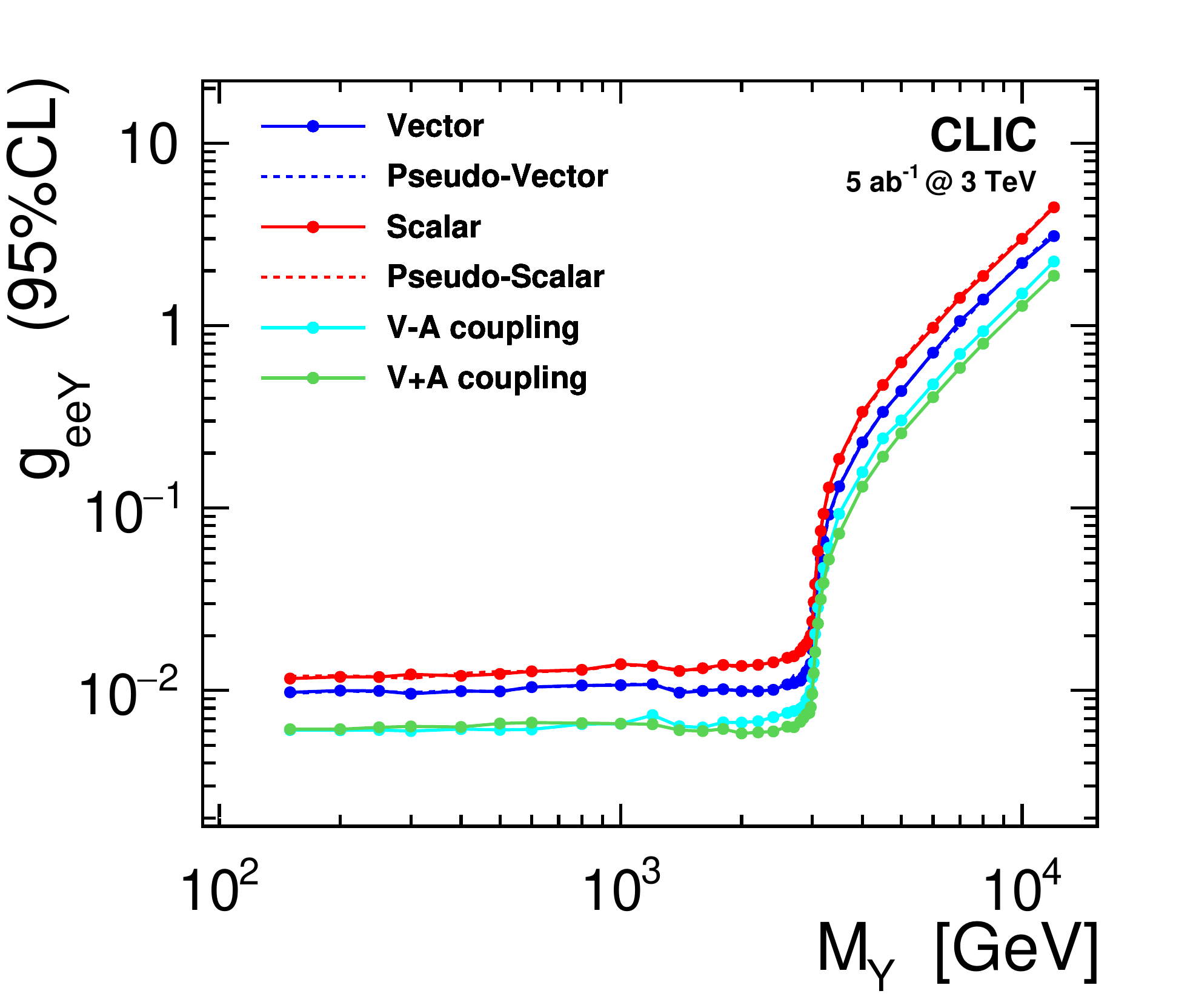}%
 }
  \caption{Limits on the mediator coupling to electrons for the
    ILC running at 500\,GeV (left) and CLIC running at 3\,TeV (right)
    for different mediator coupling scenarios and relative
    mediator width, $\Gamma/M = 0.03$. 
    Combined limits corresponding to the assumed running
    scenarios are presented with systematic uncertainties taken into
    account \cite{Kalinowski:2021tyr}.
  } 
  \label{fig:coup_wid}
\end{figure}

Shown in Fig.~\ref{fig:coup_wid} are limits on the mediator
coupling to electrons expected for different mediator coupling
scenarios and relative mediator width, $\Gamma/M = 0.03$. 
For heavy mediator exchange, coupling limits increase with the
mediator mass squared, $\mathrm{g}_\mathrm{eeY} \sim
\mathrm{M}_\mathrm{Y}^2$, as expected in the EFT limit.
Results of study \cite{Kalinowski:2021tyr} are in very good agreement
with the limits resulting from the ILD analysis
\cite{Habermehl:2020njb} based on the full detector simulation and EFT 
approach \cite{mttd2021}.

\section{Impact of polarisation}

\begin{figure}[tb]
  \centerline{
\includegraphics[width=0.4\textwidth]{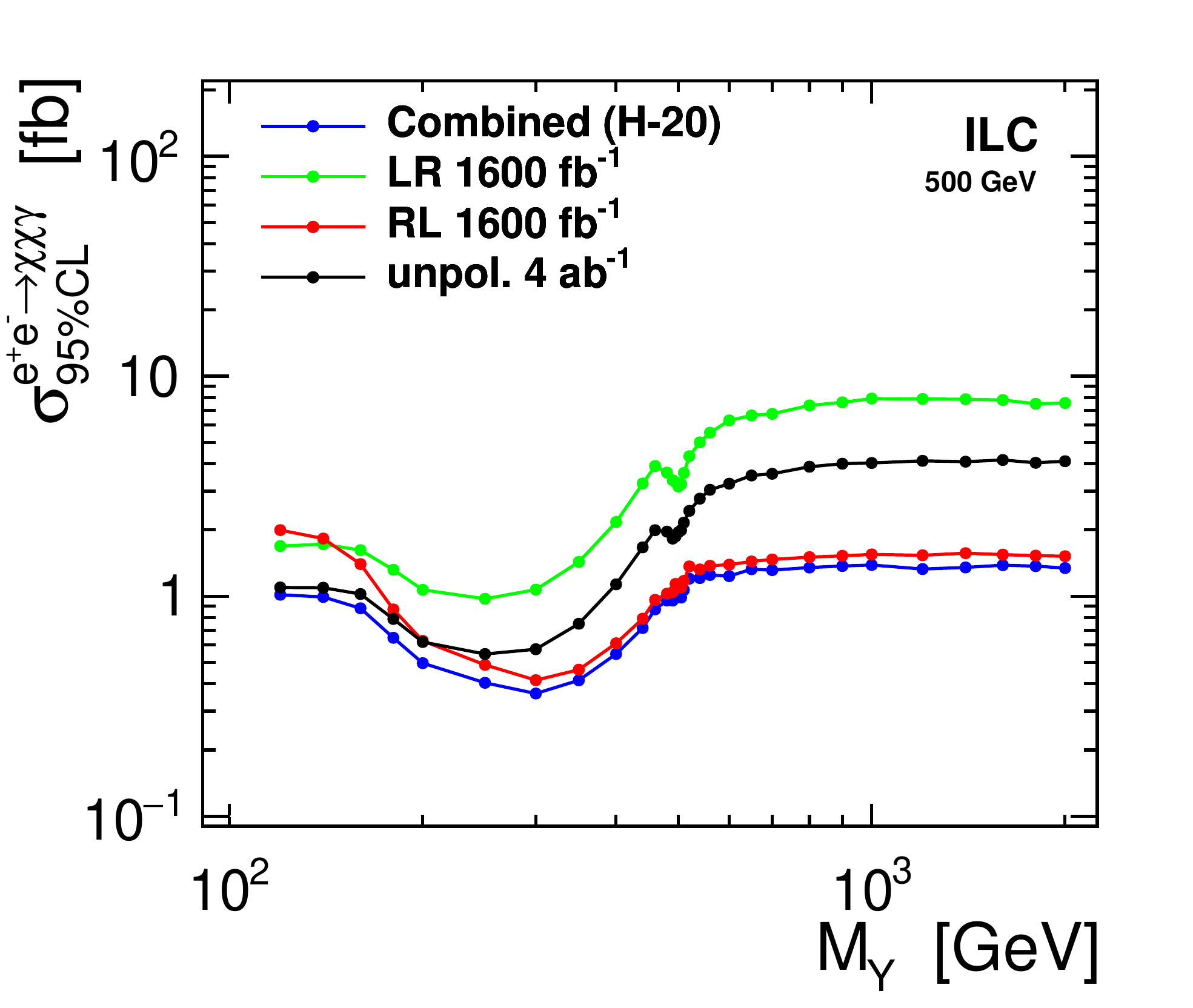}
\includegraphics[width=0.4\textwidth]{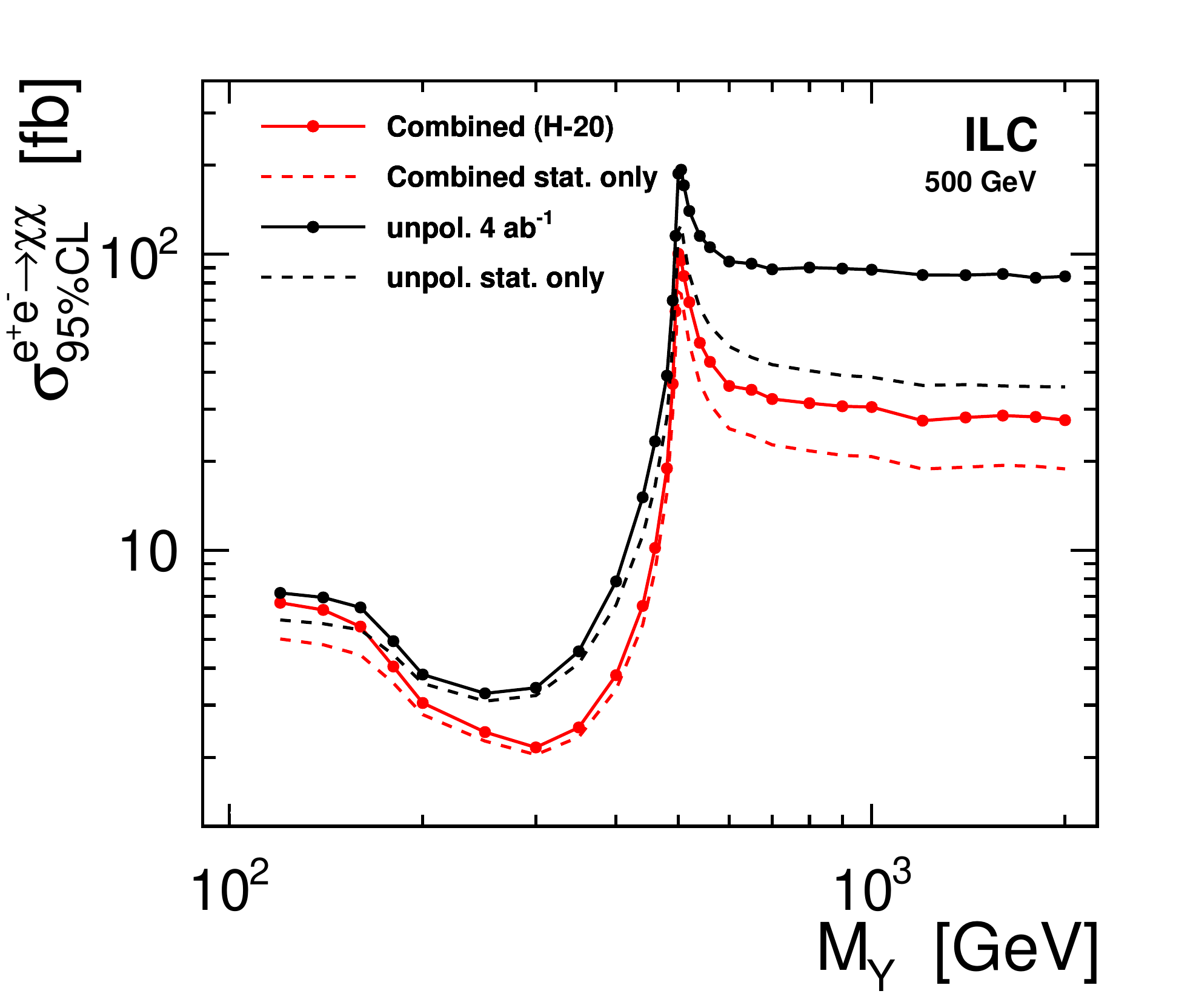}
}
\caption{Impact of beam polarisation on the expected limits 
  on the DM pair-production cross section as a function of
  vector mediator mass \cite{Kalinowski:2021tyr}.
  Left: limits from data collected with different beam polarisations
  are compared with limits from the combined data analysis and limits
  expected from unpolarised  data set with the same integrated luminosity.
  Right: limits from the combined analysis are compared with limits expected
  for unpolarised data set with (solid lines) and without (dashed
  lines) taking into account  systematic uncertainties.} 
  \label{fig:poli2}
\end{figure}

The sensitivity to processes of radiative DM production is mainly
limited by the ``irreducible'' background from radiative neutrino
pair-production events, $\epem \to \nu \,\bar{\nu} + \gamma$. 
With proper polarisation choice, this background can be strongly
suppressed and mass scale limits can improve significantly, see 
Fig.~\ref{fig:poli2} (left). 
As the structure of mediator couplings is unknown, data taken
with different polarisation combinations needs to be collected to obtain
the best sensitivity in all possible scenarios.
Moreover, by combining four independent data sets the impact of
systematic uncertainties is significantly reduced.
This is shown in Fig.~\ref{fig:poli2} (right), where the expected limits from
mono-photon analysis are compared for the combined analysis of polarised
data and for an unpolarised data set with the same integrated luminosity,
without and with systematic uncertainties taken into account.
When beam polarisation is not used, systematic uncertainties reduce
the ILC reach in EFT mass scale by almost a factor of two.
When combining data taken using different polarisation combinations,
systematic effects can be significantly constrained based on the
predicted polarisation dependence of the SM backgrounds.
For scenarios with light mediator exchange, the impact of systematic
uncertainties is reduced but 
combined analysis of data taken with different polarisation
combinations still results in significant limit improvement.

\section{Conclusions}

Future \epem colliders offer many complementary options for DM searches.
Searches based on the mono-photon signature are believed to be the most
general and least model-dependent way to look for DM production.
Dedicated procedure has been proposed for a proper simulation of
mono-photon events in \whizard \cite{Kalinowski:2020lhp} and 
the mono-photon analysis framework was developed for scenarios
with light mediator exchange and very small mediator couplings to SM
\cite{Kalinowski:2021tyr}.  
Future experiments at 500\,GeV ILC or 3\,TeV CLIC will results in
limits on the cross section for the radiative DM pair-production, 
$\epem \rightarrow \chi \chi \gamma_\mathrm{tag}$,
of the order of 1\,fb.
Limits on the mediator coupling to electrons of the order of
$\mathrm{g}_\mathrm{eeY} \sim 10^{-3} - 10^{-2}$ can be set 
up to the kinematic limit, $M_Y \le \sqrt{s}$.
For processes with light mediator exchange,
coupling limits expected from the analysis of mono-photon spectra
are stronger than those expected from the direct searches in SM decay channels.
In the heavy mediator limit,
sensitivity of future \epem colliders extends to the mediator mass
scales of the order of 10 TeV. 

\section*{Acknowledgements}

This study was supported by the National Science
Centre, Poland, the OPUS project under contract
UMO-2017/25/B/ST2/00496 (2018-2021) and the HARMONIA project under
contract UMO-2015/18/M/ST2/00518 (2016-2021), and by the German
Research Foundation (DFG) under grant number STO 876/4-1 and STO
876/2-2.

\printbibliography

\end{document}